\documentclass[reqno]{amsart}
\newtheorem{theo}{Theorem}

\newtheorem{lem}{Lemma}

\newtheorem{prop}{Proposition}

\newcommand\eps\varepsilon
\newcommand\ph\varphi
\newcommand\kap\varkappa

\renewcommand{\Re}{\mbox{\rm Re}\,}
\renewcommand{\Im}{\mbox{\rm Im}\,}

\begin{document}

\title[On analytic solutions to NSE in 3-D torus]
      {On analytic solutions to NSE in 3-D torus}

\author{Oleg Zubelevich}

\address{Department of Differential Equations\\
Moscow State Aviation Institute\\
Volokolamskoe Shosse 4, 125871, Moscow, Russia}
\email{ozubel@yandex.ru}

\date{}
\thanks{Partially supported by grants RFBR 02-01-00400, INTAS 00-221.}

\begin{abstract}We consider NSE with $H^1$-initial conditions
on the 3-dimensional torus
and
prove that there exists a solution that is analytic in all
variables. \end{abstract}

\maketitle
\numberwithin{equation}{section}
\newtheorem{theorem}{Theorem}[section]
\newtheorem{lemma}[theorem]{Lemma}
\newtheorem{definition}{Definition}[section]

\section{Introduction}

The regularity problems for NSE
have been studied by many authors.
J. Serrin
showed that if a weak solution is of the space
$L^s((0,T),L^r(D))$ with $n/r+2/s<1$, where $n$ is the dimension of
the space variables and external force is of the space
$L^1((0,T),L^1(D))$ ($D$ is a domain in $\mathbb{R}^n$ and the solution
with zero boundary conditions is considered)
then this solution is of $C^\infty$ in the spatial variables \cite{Serrin}.

Under the same assumptions C. Kahane showed that the
solution is analytic in the spatial variables \cite{Kahane}.

 It was shown by K. Masuda
 that if a solution is of $C((0,T),H^1(D))$ then this solution is as regular as
allowed by the external force (including $C^\infty$ regularity and analyticity)
\cite{Masuda}, \cite{Masuda2}.

Note that if external force and initial data are
analytic in all variables, then the (local in time)
existence of analytic solutions in all variables
immediately follows from the result announced in the end of the paper
\cite{Nishida}.

Speaking informally, all those results are about which extra
hypothesis of the weak solution is
sufficient to its regularity.

In the present paper we consider NSE on the 3-dimensional torus without external force
and show that
under $H^1$-initial conditions there exists a solution that becomes analytic in
time and spatial variables right after the beginning of the motion.

The author wishes to thank Ju. A. Dubinski\v{\i} and
A. L. Skubachevski\v{\i}  for useful discussions.

\section{Main theorem}
Consider the Navier-Stokes equation with initial data and the condition of incompressibility
of a fluid:
\begin{align}
\textbf{v}_t+(\textbf{v},\nabla) \textbf{v}&=-\nabla p+\nu \Delta
\textbf{v},\label{N-S-1}\\
\mbox{div}\,\textbf{v}&=0,\label{N-S-12}\\
\textbf{v}\mid_{t=0}&=\hat{\textbf{v}}(x),\label{N-S-11}
\end{align}
where by $\textbf{v}=(v^1(t,x),\ldots,v^3(t,x))$ we denote a vector-function,
$p(t,x)$ is a scalar function and $\nu$ is a positive constant.
The differential operators are defined in the usual way:
$\partial_j=\partial/\partial x_j$,
$$
\begin{array}{rclrcl}
\nabla p&=&(\partial_1 p,\ldots,\partial_3 p),&
\mbox{div}\,\textbf{v}&=&\sum_{k=1}^{3}\partial_k v^k,\\
\Delta f&=&\mbox{div}\,\nabla f,&
(\textbf{v},\nabla )f&=&\sum_{k=1}^{3}v^k\partial_k f.
\end{array}
$$
Application of any scalar operator to a vector-function implies
that this operator applies to each component of the vector-function.

The variables $x$ in problem (\ref{N-S-1})-(\ref{N-S-11}) belongs to the $3-$dimensional
torus: $x\in\mathbb{T}^3=\mathbb{R}^3/(2\pi\mathbb{Z})^3$.

Let $k,x\in \mathbb{C}^3$, introduce some notations:
$$
(k,x)=k_1x_1+\ldots+k_3x_3,\quad
|x|=|x_1|+\ldots+|x_3|,\quad i^2=-1.$$
Sometimes we will use the Euclidian norm:
$|x|_e^2=|x_1|^2+\ldots+|x_3|^2$, and the norm
$|x|_m=\max_{k}|x_k|$.
It is well known that there is a constant $\alpha $ such that
$$\alpha |x|_m\le|x|_e\le |x|.$$
We also
consider the following complex neighborhood of the torus:
$$\mathbb{T}^3_r=\{x\mid\Re\,x\in \mathbb{T}^3,\quad|\Im\,x|_m<r\}.$$

Due to the Galilean invariance of the equations (\ref{N-S-1})-(\ref{N-S-11})
(they are not changed after the substitution
$x\mapsto x+\textbf{c}t,\quad \textbf{v}\mapsto \textbf{v}+\textbf{c}$,
where $\textbf{c}$ is a constant) we
assume without loss of generality that
the initial vector field $\hat{\textbf{v}}(x)$ is of zero mean value:
$$
\int_{\mathbb{T}^3}\hat{\textbf{v}}(x)\,dx=0.$$
All the functions below are of zero mean value.
So, define the set $\mathbb{Z}^3_0=\mathbb{Z}^3\backslash\{0\},$
and
denote by $H^s(\mathbb{T}^3)$ the Sobolev space of zero mean value
functions with the norm:
$$f(x)=\sum_{k\in\mathbb{Z}^3_0}f_ke^{i(k,x)},\quad
\|f\|^2_{H^s(\mathbb{T}^3)}=\sum_{k\in\mathbb{Z}^3_0}|f_k|^2|k|_e^{2s}.$$

Drop the subscript for the $H^1(\mathbb{T}^3)$-norm:
$$\|\cdot\|=\|\cdot\|_{H^1(\mathbb{T}^3)}.$$

Let $I_T=[0,T]$.
We use the norm
$$\|u(t,x)\|^C=\sup_{t\in I_T}\|u(t,\cdot)\|$$
in $C(I_T,H^1(\mathbb{T}^3))$.

The substitution $p\mapsto p+c(t)$ ($c(t)$ is an arbitrary function)
in (\ref{N-S-1}) also does not change the
equation. So we will find the function $p$ just up to an additional function
of $t$.

Define the set
$$\begin{aligned}
Q_r(T)=&
\{t\in \mathbb{C}\mid |\Im t|<\Re t<T,\quad r<\Re t\}\times
\mathbb{T}^3_{\frac{r\nu\alpha }{4}}\\
=&\Big\{(t,x)\mid |\Im t|<\Re t<T,\quad \frac{4}{\nu\alpha }|\Im x|_m<r<\Re t,
\quad\Re x\in\mathbb{T}^3\Big\}.\end{aligned}$$
The set $Q_r(T)$ is open as a cross product of open sets.
The set
$$
\begin{aligned}
Q (T)=&\bigcup_{0<r<T}Q_r(T)\\=&
\Big\{(t,x)\mid |\Im t|<\Re t<T,\quad \frac{4}{\nu\alpha }|\Im x|_m<\Re t,
\quad\Re x\in\mathbb{T}^3\Big\}\end{aligned}$$
is open as a union of open sets.

Assume that
the initial vector field $\hat{\textbf{v}}(x)$ belongs to
the space $ H^1(\mathbb{T}^3)$ and
$\mbox{div}\,\hat{\textbf{v}}(x)=0.$ We mean that the operator
$\mbox{div}$ consists of generalized derivatives. Namely,
for any function $f(x)=\sum_{k\in \mathbb{Z}^3}f_ke^{i(k,x)}\in
L^2(\mathbb{T}^3)$ the series
$$f_{x_j}(x)=\sum_{k\in \mathbb{Z}^3}ik_jf_ke^{i(k,x)}$$
possibly diverges but it defines a generalized function which
value of a test-function $\phi(x)\in C^\infty(\mathbb{T}^3)$ computes as
follows:
$$(f_{x_j},\phi)=\int_{\mathbb{T}^3}f_{x_j}(x)\phi(x)\,dx=
\sum_{k\in \mathbb{Z}^3}ik_jf_k\int_{\mathbb{T}^3}e^{i(k,x)}\phi(x)\,dx.$$

Denote by $\mathcal{O}(D)$ the space of holomorphic functions in a domain
$D$.

\begin{theo}
\label{main_theo}There exists a positive constant $c$ that depends only on
$\nu$ such that for $$T=\frac{c}{(1+\|\hat{\mathbf{v}}\|)^{16}}$$
\begin{enumerate}\item\label{st1}
initial problem (\ref{N-S-1})-(\ref{N-S-11}) has a solution:
$$\mathbf{v}(t,x)=\sum_{k\in
\mathbb{Z}^3_0}\mathbf{v}_k(t)e^{i(k,x)}\in\mathcal{O}(Q(T)) ,$$
and
$$\|\hat{\mathbf{v}}-\mathbf{v}(t,\cdot)\|\to 0\quad
\mathrm{as}\quad t\to 0.$$
Furthermore, there is a function
$$V(t,x)=\sum_{k\in
\mathbb{Z}^3_0}V_k(t)e^{i(k,x)}\in C(I_T, H^1(\mathbb{T}^3)),$$
such that for $|\Im\, t|<\Re\,t$ we have:
$$|\mathbf{v}_k(t)|\le V_k(\Re t)\exp(-\Re t|k|_e\nu/2).$$

\item\label{st2} There exists a positive  constant $\mu$ such that from
 $\|\hat{\mathbf{v}}\|\le\mu$ it follows that the first part of this theorem remains valid
for $T=\infty$.
\end{enumerate}
\end{theo}

As a corollary we have:
\begin{prop}Under the conditions of the second part of the theorem
for any $t\ge 0$
we have an estimate
$$\|\mathbf{v}(t,x)\|\le ce^{-\nu t/2},$$
a positive constant $c$ does not depend on $t$.
\end{prop}

As it was established by Leray \cite{Leray_9}
there exits a solution which  belongs to
$$L^2(\mathbb{R}_+,H^1(\mathbb{T}^3))
\bigcap L^\infty(\mathbb{R}_+,L^2(\mathbb{T}^3)),$$
so called weak solution.

On the other hand by Serrin's result \cite{Serrin_12}
a regular solution, as long as it exists, is unique
in the class of weak solutions.

Both of these theorems in conjunction with Theorem \ref{main_theo},
involve that the weak solution
is analytic in all variables except the set of such singular values $\tilde t$ for
which $\lim_{t\to \tilde t-0}\|\mathbf{v}(t,\cdot)\|=\infty$.

This set has been studied by many
authors but nobody still knows whether it is empty or not. For example,
Leray studied the possible occurrence of singularities and noticed that
this set has Lebesgue measure $0$ and even a $1/2$-Hausdorff dimension $0$
in $[0,T]$. Furthermore the complement of this set in  $[0,T]$ is a
countable union of semiclosed intervals $[a_i,b_i)$. In the special case
we discuss, these results follow from Theorem \ref{main_theo}.

As another consequence of Theorem \ref{main_theo} note that if a weak
solution $\mathbf{v}(t,x)$ is equals to zero at a moment $t$ for
infinite number of points $x\in \mathbb{T}^3$
then it equals to zero identically at least for all $t$
between two singularities.

Before approaching to a proof of the theorem we must develop
some technique tools.

\section{Definitions and technique tools}
We will denote inessential
constants by $c,C$ or by these letters with subscripts.

Provide the space $\mathcal{O}(D)$ with a collection of norms: let $u\in
\mathcal{O}(D)$ and $K$ be a subset of $D$, then
$\|u\|_K=\sup_{z\in K}|u(z)|$.
A sequence $u_k\in \mathcal{O}(D)$ converges to $u\in
\mathcal{O}(D)$ if for any compact  $K$ we have $\|u_k-u\|_K\to 0$ when $k\to
\infty$. This kind of convergence is referred to as compact convergence.
Equipped with such a convergence the space $\mathcal{O}(D)$ becomes a
seminormed space.

The continuity of a function of $\mathcal{O}(D)$ implies
the continuity with respect to the compact convergence.

Particularly, a linear operator $A:\mathcal{O}(D)\to \mathcal{O}(D)$ is
continues iff for any compact set $K\subset D$ there exists a compact set
$K'\subset D$ and a constant $c$ such that for all $u\in\mathcal{O}(D)$
we have $\|Au\|_K\le
c\|u\|_{K'}$ and the constant $c$ does not depend on $u$.

Evidently, the compact convergence in the space $\mathcal{O}(Q(T))$
follows from the convergence with respect to the norms
$\|\cdot\|_{Q_r(T)}$.

We say that a subset $M$ of $\mathcal{O}(D)$ is bounded if for any compact
set
$K\subset D$ there is a constant
$C_K$ such that if $u$ belongs to $M$ then $\|u\|_K\le C_K$.

Other details about seminormed spaces topology there are in
\cite{Schwartz}, \cite{Yosida_mon}.

\begin{theo}[Montel]
\label{Montel}
If $M$ is a closed and bounded subset of $\mathcal{O}(D)$, then it is a
compact set.
\end{theo}
The following theorem is a spetial case of the result of \cite{Browder}.
\begin{theo}
\label{main_t}Let $K$ be a convex compact subset of $\mathcal{O}(D)$.
Then a continues map $f:K\to K$ has a fixed point $\hat {x}\in K$ i.e.
$f(\hat {x})=\hat {x}$.
\end{theo}

The Laplace operator $\Delta$ is invertible: there exists a bounded operator
$\Delta^{-1}:L^2(\mathbb{T}^3)\to
L^2(\mathbb{T}^3)$ such that if
$$
u(x)=\sum_{k\in\mathbb{Z}^3_0}u_ke^{i(k,x)}$$ is the
Fourier expansion of $u$ then an explicit
form of this operator is
\begin{equation}
\label{oblap}
\Delta^{-1}u(x)=-\sum_{k\in\mathbb{Z}^3_0}\frac{u_k}{|k|_e^2}e^{i(k,x)}.
\end{equation}

Formula (\ref{oblap}) involves that for any function $f\in\mathcal{O}(\mathbb{T}_R^3)$
there is an estimate:
$$\|\Delta^{-1}f\|_{\mathbb{T}_r^3}\le
c(r,\delta)\|f\|_{\mathbb{T}_{r+\delta}^3},\quad \delta>0,\quad r+\delta<R.$$
So, $\Delta^{-1}$ is a continues operator of the space
$\mathcal{O}(\mathbb{T}_R^3)$ to itself.

Let $$u(x)=\sum_{k\in \mathbb{Z}^3_0}u_ke^{i(k,x)}\in L^2(\mathbb{T}^3).$$
Define semigroups by the formulas:
$$\begin{aligned}
S^{t}u(x)&=\sum_{k\in\mathbb{Z}^3_0}u_ke^{i(k,x)-|k|^2_et},\\
R^tu(x)&=\sum_{k\in\mathbb{Z}^3_0}u_ke^{i(k,x)-|k|_et/2},\\
H^t&=S^tR^{-t}.\end{aligned}.$$
For any fixed $t\ge 0$ the operators $S^t, R^t,H^t$ are continues on
$\mathcal{O}(\mathbb{T}_R^3)$.

\subsection{Majorant functions}\label{majjjj}
Let
$$
\begin{aligned}
v(t,x)&=\sum_{k\in\mathbb{Z}_0^3}v_k(t)e^{i(k,x)}\in  \mathcal{O}(Q(T)),\\
V(\tau,x)&=\sum_{k\in\mathbb{Z}^3_0}V_k(\tau)e^{i(k,x)}\in
C(I_T, H^1(\mathbb{T}^3)).\end{aligned}
$$
A notation $v\ll V$
means that  $|v_k(t)|\le V_k(\Re t)$ holds for all $|\Im\,t|<\Re\,t<T$ and
$k\in \mathbb{Z}^3_0$.

If $u,U$ are vector-functions, then a relation $u\ll U$ means that each component of
the vector $U$ majorates corresponding component of the vector $u$.

Define a continues linear operator of $\mathcal{O}(\mathbb{T}^3_R)$ to itself:
$$ Du=\sum_{k\in \mathbb{Z}^3_0}|k|_eu_ke^{i(k,x)}.$$
The operator $-D$ is an infinitesimal operator for the semigroup $R^{2t}$
and the operator $\Delta$ is an infinitesimal operator for the semigroup
$S^t$.

Define the contour $L(t)$ as follows:
$$
L(t)=\left\{s+i\frac{\Im t}{\Re t}s\mid 0\le s\le \Re t,\right\},\quad
|\Im\,t|<\Re\,t<T.
$$

Enumerate main properties of the relation "$\ll$".  Suppose
$u(t,x)\ll U(\tau,x)$ and $v(t,x)\ll V(\tau,x)$; then:
$$\begin{array}{rclrcl}
u+v&\ll&U+V,&uv&\ll&UV,\\
\lambda u&\ll& |\lambda|U,&\int\limits_{L(t)}u(s,x)\,ds&\ll&
2\int\limits_{0}^{\mathrm{Re}\, t}U(s,x)\,ds,\\
\partial_l u&\ll& DU,& S^{\nu t}u&\ll& U,\\
D(uv)&\ll&UDV+VDU,& \Delta^{-1}u&\ll& U.
\end{array}$$
In these formulas $\lambda$ is a complex number.

Another property of "$\ll$" is as follows: there exists some positive
constant $c$ such that an estimate
\begin{equation}
\label{ewwrwr}
\Delta^{-1}\partial_j\partial_l u\ll c\,U
\end{equation}
holds for all functions $u\ll U$. Indeed,
expanding the left- and the
right-hand  side of (\ref{ewwrwr}) to Fourier series by formula (\ref{oblap})
we, see
that the estimate follows from the inequality:
$$|k_j k_l|\le c|k|_e^2,\quad k\in \mathbb{Z}^3_0.$$

 We say that a map $F:C(I_T, H^1(\mathbb{T}^3))\to C(I_T, H^1(\mathbb{T}^3))$
 majorates a map $f:\mathcal{O}(Q(T))\to \mathcal{O}(Q(T))$
  (denote by $f\ll F$) if for any functions $u,U$
the relation $u\ll U$ involves $f(u)\ll F(U)$.

\begin{lem}
\label{m_kor}
Let $u,v\in H^1(\mathbb{T}^3)$ be of zero mean functions, then
$$\|H^tD(u\cdot v)\|\le h(t)\|u\|\cdot\|v\|,\quad
h(t)=\left\{
\begin{aligned}
\frac{c}{t^{7/8}}&\quad{\rm if}& 0<t\le 1,\\
ce^{-t/2} &\quad{\rm if}& t>1,
\end{aligned}
\right.
$$positive constant $c$ does not depend on $u$ and $v$.
(Note that the product $u\cdot v$ belongs at least to
$L^1(\mathbb{T}^3)$.)
\end{lem}
{\it Proof.} Consider the estimate for $0<t\le 1$. The estimate for $t>1$
is easy to obtain.
Introduce some notations:
$$
\begin{aligned}
J_k=&\{(m,n)\in \mathbb{Z}^3_0\times\mathbb{Z}^3_0\mid
m+n=k\ne 0\},\\
N_k=&\{(m,n)\in J_k\mid |m|_e\ge |k|_e,\quad |n|_e\ge |k|_e\},\\
M_k=&J_k\backslash N_k.
\end{aligned}
$$
The set $M_k$ is finite and every pair $(m,n)\in M_k$ estimates as
follows:
\begin{equation}
\label{n2k}
|m|_e< 2|k|_e,\quad |n|_e< 2|k|_e.\end{equation}
Indeed, if $|n|_e\ge |k|_e$ then $|m|_e<|k|_e$ and by the equality $n=k-m$
we have $|n|_e\le |k|_e+|m|_e< 2|k|_e$.

Let
$$u(x)=\sum_{k\in \mathbb{Z}^3_0}u_ke^{i(k,x)},\quad
v(x)=\sum_{k\in \mathbb{Z}^3_0}v_ke^{i(k,x)},$$
then
$$H^tDu=\sum_{k\in \mathbb{Z}^3_0}|k|_e
e^{-|k|^2_et(1-1/(2|k|_e))+i(k,x)}u_k,$$
and
$$H^tD(u(x)\cdot v(x))=
\sum_{k\in \mathbb{Z}^3_0}
|k|_ee^{-|k|^2_et(1-1/(2|k|_e))+i(k,x)}
\sum_{(m,n)\in J_k}u_n\cdot v_m.$$

So,
by the formula
$$
e^{-|k|^2_et(1-1/(2|k|_e))}\le e^{-|k|^2_et/2}$$
we have:
\begin{equation}
\label{man}\|H^tD(u\cdot v)\|^2\le
\sum_{k\in \mathbb{Z}^3_0}e^{-|k|^2_et}\cdot|k|_e^4\cdot
\Big(\sum_{(m,n)\in J_k}|u_n|\cdot|v_m|\Big)^2.\end{equation}
Substitute to  (\ref{man}) the expansion:

\begin{align}
\Big(\sum_{(m,n)\in J_k}|u_n|\cdot|v_m|\Big)^2=
\Big(\sum_{(m,n)\in N_k}|u_n|\cdot|v_m|\Big)^2+&\nonumber\\
2\sum_{(m,n)\in N_k}|u_n|\cdot|v_m|\cdot\sum_{(m,n)\in
M_k}|u_n|\cdot|v_m|+&
\Big(\sum_{(m,n)\in M_k}|u_n|\cdot|v_m|\Big)^2,\label{maiin}
\end{align}
and estimate these sums separately. By the Cauchy inequality one has:
\begin{align}
\sum_{(m,n)\in N_k}|u_n|\cdot|v_m|\le
\frac{1}{|k|_e^2}
\sum_{(m,n)\in N_k}|n|_e|u_n|\cdot|m|_e|v_m|\le\nonumber\\
\frac{1}{|k|_e^2}\sqrt{\sum_{n\in\mathbb{Z}^3_0}|n|_e^2|u_n|^2}
\sqrt{\sum_{n\in\mathbb{Z}^3_0}|n|_e^2|v_n|^2}=
\frac{1}{|k|_e^2}\|u\|\cdot\|v\|\label{o_ne}.
\end{align}

Thus for the first sum of (\ref{man})-(\ref{maiin}) we have:
\begin{align}
\sum_{k\in \mathbb{Z}^3_0}e^{-|k|^2_et}|k|_e^4
\Big(\sum_{(m,n)\in N_k}|u_n|\cdot|v_m|\Big)^2&\le
\|u\|^2\|v\|^2\sum_{k\in \mathbb{Z}^3_0}e^{-|k|^2_et}\nonumber\\&\le
\frac{c}{t^{3/2}}\|u\|^2\|v\|^2.\label{uw}\end{align}
The last inequality follows from the formula
\begin{equation}
\label{qi}
\sum_{k\in \mathbb{Z}^3_0}e^{-|k|^2_et}\le
c\Big(\int_0^\infty e^{-x^2t}\,dx\Big)^3\le
\frac{c}{t^{3/2}}.\end{equation}

Denote the last sum of (\ref{man})-(\ref{maiin})
by $A$ and estimate it with the help of formulas (\ref{n2k}):
$$
\begin{aligned}
A=&\sum_{k\in \mathbb{Z}^3_0}e^{-|k|^2_et}|k|_e^4
\Big(\sum_{(m,n)\in M_k}|u_n|\cdot|v_m|\Big)^2\\
\le&\sum_{k\in \mathbb{Z}^3_0}e^{-|k|^2_et/2}|k|_e^{\frac{1}{2}}
|k|_e^{\frac{7}{2}}
\Big(\sum_{(m,n)\in
M_k}e^{-\frac{|n|^2_et}{32}}|u_n|e^{-\frac{|m|^2_et}{32}}|v_m|\Big)^2.
\end{aligned}
$$
Using the inequality
$$e^{-|k|^2_et/2}|k|_e^{\frac{1}{2}}\le \frac{c}{t^{1/4}},\quad k\in \mathbb{Z}^3_0$$
we obtain
$$A\le\frac{c}{t^{1/4}}\|S^{t/32}u\cdot S^{t/32}v\|^2_{H^{7/4}(\mathbb{T}^3)}\le
\frac{c}{t^{1/4}}\|S^{t/32}u\|^2_{H^{7/4}(\mathbb{T}^3)}\|S^{t/32}v\|^2_{H^{7/4}(\mathbb{T}^3)}.
$$
Estimate the term:
$$\|S^{t/32}u\|^2_{H^{7/4}(\mathbb{T}^3)}=
\sum_{k\in \mathbb{Z}^3_0}
e^{-\frac{|k|^2_et}{32}}|k|_e^{3/2}|k|_e^2|u_k|^2\le
\frac{c}{t^{3/4}}\|u\|^2,$$
we use again the  inequality
$$e^{-|k|^2_et/32}|k|_e^{\frac{3}{2}}\le \frac{c}{t^{3/4}},\quad k\in \mathbb{Z}^3_0.$$
Finally
it follows that
\begin{equation}
\label{mui}
A\le\frac{c}{t^{7/4}}\|u\|^2\|v\|^2.\end{equation}

Estimate the middle sum of (\ref{man})-(\ref{maiin}) by  formula (\ref{o_ne}):
$$
\begin{aligned}
B&=\sum_{k\in \mathbb{Z}^3_0}e^{-|k|^2_et}|k|_e^4
\sum_{(m,n)\in N_k}|u_n|\cdot|v_m|\cdot\sum_{(m,n)\in
M_k}|u_n|\cdot|v_m|\\ &\le
\|u\|\cdot\|v\|\sum_{k\in \mathbb{Z}^3_0}e^{-|k|^2_et}|k|_e^2
\sum_{(m,n)\in M_k}|u_n|\cdot|v_m|.
\end{aligned}
$$
Then using the Cauchy inequality and formulas (\ref{qi}), (\ref{mui})
one has:
$$
\begin{aligned}
B&\le \|u\|\cdot\|v\|
\sqrt{\sum_{k\in \mathbb{Z}^3_0}e^{-|k|^2_et}}
\sqrt{\sum_{k\in \mathbb{Z}^3_0}e^{-|k|^2_et}|k|_e^4(\sum_{(m,n)\in
M_k}|u_n|\cdot|v_m|)^2}\\&\le
\frac{c}{t^{13/8}}\|u\|^2\|v\|^2.
\end{aligned}$$
Gathering this estimate with (\ref{uw}) and (\ref{mui})
we obtain the assertion of the Lemma.

Lemma is proved.

\subsection{Existence Lemma}
Define a map
$$\Phi(u,v)=\int_0^tH^{\nu(t-\xi)}D(u(\xi,x)v(\xi,x))\,d\xi.$$
\begin{lem}
\label{first_lem}
The map $\Phi$ takes the space $C(I_T, H^1(\mathbb{T}^3))\times
C(I_T, H^1(\mathbb{T}^3))$ to the space $C(I_T, H^1(\mathbb{T}^3)),$
and
\begin{equation}
\label{est_firs}
\|\Phi(u,v)\|^{C}\le c\|u\|^{C}\|v\|^{C},
\end{equation}
positive constant $c$ does not depend on $u,v$ and $T$.
\end{lem}
{\it Proof.} Taking $\tau=t-\xi$, by Lemma \ref{m_kor} we have:
$$
\|\Phi(u,v)\|^{C}=\|u\|^C\|v\|^C\sup_{t\ge 0}\int_{0}^t h({\nu}\tau)\,d\tau.
$$Evidently, the last $\sup$ is lower than infinity.
Lemma is proved.

Define a function $\hat V(x)$ as follows:
\begin{equation}
\label{lkjhg}
\hat V(x)=\sum_{k\in\mathbb{Z}^3}|\hat{\mathbf{v}}_k|_me^{i(k,x)},\end{equation}
where $\hat{\mathbf{v}}_k$ is the Fourier coefficients of the initial data
$\hat{\mathbf{v}}(x)$. Evidently we have
$\|\hat V\|=\|\hat{\mathbf{v}}\|$.
Let $$a=\frac{2}{\alpha }.$$
\begin{lem}
\label{exiseee}
Let the
function $\hat V\in  H^1(\mathbb{T}^3)$ is defined by (\ref{lkjhg})
then
 the equation
\begin{equation}
\label{ex_eqr}
V(t,x)=H^{\nu t}\hat V(x)+a\int_0^tH^{(t-\xi)\nu}D(V^2)(\xi,x)\,d\xi
\end{equation}
 has a unique solution $V(t,x)\in C(I_T, H^1(\mathbb{T}^3))$
  with the same constant $T$ as in Theorem \ref{main_theo}.

If $\sup_{t\ge 0}\|H^{\nu t}\hat V(x)\|$
 is sufficiently small then previous assertion remains valid for
$T=\infty$.
 \end{lem}

{\it Proof.} We shall prove that for sufficiently small $T$ the map
$$F(V)=H^{\nu t}\hat V(x)+a\int_0^tH^{(t-\xi)\nu}D(V^2)(\xi,x)\,d\xi$$
 is a contraction of the ball
$$B=\{V\in C(I_T, H^1(\mathbb{T}^3))\mid \|V-H^{\nu t}\hat
V(x)\|^C\le 1\}.$$
For selected $T$ a parameter
$$\eps=\sup_{t\in I_T}\int_{0}^t h(\nu\tau)\,d\tau=cT^{1/8}$$
is sufficiently small to operator $F$ be contraction of the ball $B$.
Indeed, let $V\in B$ it follows that
$$\|V\|^C\le 1+\|H^{\nu t}\hat V\|^C$$
and take $T$ such small as $$
\|F(V)-H^{\nu t}\hat V\|^C\le a\eps(\|V\|^C)^2\le 1$$
thus $F(B)\subseteq B$.
Let $V',V''\in B$ then
$$\|F(V')-F(V'')\|^C\le a\eps\|V'+V''\|^C\cdot\|V'-V''\|^C.$$
One can check that
$$a\eps\|V'+V''\|^C<1$$ so $F$ is contraction of the ball
$B$.

The second assertion of the lemma follows from similar arguments.

Lemma is proved.

\section{Proof of theorem \protect\ref{main_theo}}
We shall prove Theorem \ref{main_theo} by the Majorant functions method. Namely,
the original problem will be replaced with the so called
majorant problem. Then we prove the existence theorem for the
majorant problem and show that it involves the existence theorem for the
original one.

The majorant functions method was originated by Cauchy and
Weierstrass and applied by Kovalevskaya
to prove an existence of analytic solutions to initial problems for PDE.
Further studies and applications of this technique contains in
\cite{lednev},\cite{Treshev3}, \cite{zu},  \cite{zu2}.

Now prove the first part of the theorem, the second one results in the same
way due to the last assertion of Lemma \ref{exiseee}.

Apply formally some standard procedure.
The permissibility of such a procedure  will be clear from the
further context.

So,
take the operator $\mbox{div}$ from the right- and the left-hand sides of
equation (\ref{N-S-1}). Using equation (\ref{N-S-12}) we get
$\partial_i\partial_j (v^iv^j)=-\Delta p,$
where we summarize by the
repeated subscripts. Thus
\begin{equation}
\label{pres}
p=-\Delta^{-1}\partial_i\partial_j (v^iv^j).\end{equation}
Substituting this formula to equation (\ref{N-S-1}) we obtain the
following problem:
\begin{equation}
\label{N-S_main}
\begin{aligned}
(v^k)_t&=A^k_l\partial_j(v^jv^l)+\nu\Delta
v^k,\quad A^k_l=(\Delta^{-1}\partial_k\partial_l-\delta_{kl}),\\
v^k\mid_{t=0}&=\hat{v}^k,
\end{aligned}
\end{equation}
where $\delta_{kl}=1$ for $k=l$ and $0$ otherwise.

The operator $A^k_l$ is continues on the space $\mathcal{O}(\mathbb{T}^3_R)$:
\begin{equation}
\label{yyz}
\begin{aligned}
\|A^k_lu\|_{\mathbb{T}^3_r}&\le c_{r,\delta}
\|u\|_{\mathbb{T}^3_{r+\delta}}.
\end{aligned}
\end{equation}

Present equation (\ref{N-S_main}) in
the form:
\begin{equation}
\label{plmnjk}
\begin{aligned}
(v^k)(t,x)&=G^k(\mathbf{v})=S^{\nu t}\hat
v^k(x)+\int_{L(t)}S^{\nu(t-\xi)}A^k_l\partial_j(v^jv^l)(\xi,x)\,d\xi,\\
G(\mathbf{v})&=(G^1,\ldots, G^3).
\end{aligned}
\end{equation}

\begin{lem}Let the function $\hat V$ is defined by (\ref{lkjhg})
and the function $V(t,x)$ is the solution of equation (\ref{ex_eqr})
 then
the map $G$ takes the  set
$$W=\{u\in \mathcal{O}(Q(T))\mid u\ll
R^{\nu\mathrm{Re}\, t}V,\quad\mathrm{div}\, u=0\}$$
into itself.
\end{lem}

{\it Proof.}
It is easy to check that the map $G$ takes a solenoidal vector-field to
a solenoidal vector-field, indeed it follows from the equality:
$$\partial_k A_l^k=0.$$

Due to the choice of the constant $a$  for any
$$v^k\ll R^{\nu\mathrm{Re}\,t}V$$ we have
 $$G^k(\mathbf{v})\ll S^{\nu\mathrm{Re}\,t}
  \hat V+a\int_0^{\mathrm{Re}\,t}S^{\nu(\mathrm{Re}\,t-\xi)}
 D(R^{\nu\xi}V)^2\,d\xi.$$
Then using the inequality $|j+k|_e\le |j|_e+|k|_e$ one check that
$$(R^tV)^2\ll R^{t}(V)^2.$$
Thus we have
 $$\begin{aligned}
S^{\nu\tau}
  \hat V&+a\int_0^{\tau}S^{\nu(\tau-\xi)}
 D(R^{\nu\xi}V)^2\,d\xi\\ &\ll R^{\nu\tau}
 \Big(H^{\nu\tau}\hat V+a\int_0^{\tau}H^{\nu(\tau-\xi)}
 D(V)^2\,d\xi\Big)=R^{\nu\tau} V.\end{aligned}$$

Lemma is proved.

\begin{lem}
\label{gggggk}The set $W$ is convex and it is compact in $\mathcal{O}(Q(T))$.
\end{lem}
{\it Proof.} The convexity is obvious.

According to the Montel theorem it is sufficient to prove that the set $W$ is
bounded. For $(t,x)\in Q_r(T)$ by the estimate $|e^{i(j,x)}|\le e^{|j||\Im\,x|_m}$ we have
$$
\begin{aligned}
|v^k(t,x)|&\le\Big|\sum_{j\in\mathbb{Z}^3_0}v^k_j(t)e^{i(j,x)}\Big|\le
\sum_{j\in\mathbb{Z}^3_0}|v^k_j(t)|e^{|j|\nu \alpha r/4}\\&\le
\sum_{j\in\mathbb{Z}^3_0}V_j(\Re t)e^{-\nu|j|_er/2+|j|\nu \alpha  r/4}
\le\sum_{j\in\mathbb{Z}^3_0}V_j(\Re t)e^{-\nu r|j|_e/4}.
\end{aligned}
$$
Thus \begin{equation}
\label{po}\|v^k\|_{Q_r(T)}\le
\sup_{t\in I_T}\sum_{j\in\mathbb{Z}^3_0}
V_j(\Re t)e^{-\nu r|j|_e/4}.\end{equation}
The right-hand side of  estimate (\ref{po}) is bounded for all
$r>0$.
It follows from the fact that $V\in
C(I_T, H^1(\mathbb{T}^3)).$

Lemma is proved.

The map $G$ is continuous with respect to the seminormed topology in
$\mathcal{O}(Q(T))$. Thus according to the Theorem \ref{main_t} and
Lemma \ref{gggggk}
 it has a fixed point $\mathbf{v}(t,x)\in W$.  This fixed point is
 a solution of equations (\ref{N-S-1}), (\ref{N-S-12}) for $(t,x)\in Q(T)$.

The Theorem is proved.


\begin{thebibliography}{99}
\bibitem{Browder}     F. E. Browder A new generalization of the Schauder
                      fixed point theorem, Math. Ann. 174, (1967), 285-290.



\bibitem{Kahane}      C. Kahane On the spatial analyticity of solutions of
the Navier-Stokes equations. Arch. Rational Mech. Anal. 33 (1969) p.
386-405.


\bibitem{lednev}    N. Lednev New method for solving PDE. Math. Collection
                        22 (64) 1948 P. 205-259  (in Russion)
\bibitem{Leray_9}  J. Leray Essai sur le mouvement d'un liquide visqueux
emplissant l'espace, Acta Math. 63 (1934), 193-248



\bibitem{Masuda}   K. Masuda On the analyticity and the unique
continuation theorem for solutions of the Navier-Stokes equations. Proc.
Japan Acad. 43 (1967) p. 827-832.


\bibitem{Masuda2} K.Masuda, On the regularity of solutions of the nonstationary
Navier-Stokes equations,
Approx. methods for N.-Stokes problems, Lect. Notes in Math., 771
(1980), 360-370



\bibitem{Nishida}     T. Nishida A Note On A Theorem Of Nirenberg
J. Differential Geometry 12 (1977) 629-633.


\bibitem{Treshev3}    A. Pronin, D. Treschev Continuous averaging in
                      multi-frequency slow-fast systems.
                      Regular and Chaotic Dynamics, V.5, No.2, 2000,
                        157--170.


\bibitem{Serrin_12} J. Serrin The initial value problem for the
Navier-Stokes equations, Non-linear Problems R. E. Langer, ed., Univ. of
Wisconsin Press, 1963, pp. 69-98.

\bibitem{Serrin}   J. Serrin On the interior regularity of weak solutions
of the Navier-Stokes equations.
Arch. Rational Mech. Anal. 9 (1962) p.
187-195.

\bibitem{Schwartz} L. Schwartz Analyse Math$\acute{e}$matique, Hermann, 1967.


\bibitem{Yosida_mon} K. Yosida Functional Analysis, Springer Verlag 1965.

\bibitem{zu}         O. Zubelevich On the Magorant Method
                     for Cauchy-Kovalevskaya Problem.
                     Mathematical Notes, 2001, 69(3), 363-374.
                                          (in Russian)

\bibitem{zu2}   O. Zubelevich, A generalization of Schauder's theorem
  and its application to Cauchy-Kovalevskaya problem,
  Electron. J. Diff. Eqns., Vol. 2003(2003), No. 55,
  pp. 1-6.



\end{thebibliography}
 \end{document}